\documentclass[aip,amsmath,amssymb,reprint]{revtex4-1}

\usepackage{graphicx}
\usepackage{dcolumn}
\usepackage{amssymb}
\usepackage{amsmath}

\usepackage{hyperref}

\usepackage[utf8]{inputenc}
\usepackage[T1]{fontenc}
\usepackage{mathptmx}
\usepackage{etoolbox}

\newcommand{\nn}{\nonumber}
\newcommand{\vphi}{\varphi}
\newcommand{\veps}{\varepsilon}
\newcommand{\bea}{\begin{eqnarray}}
\newcommand{\eea}{\end{eqnarray}}

\makeatother

\begin{document}

\title[]{Phase lines in mean-field models with nonuniform external forces}
\author{Roni Kroll and Yoav Tsori}
\email{tsori@bgu.ac.il}
\affiliation{Department of Chemical Engineering, Ben-Gurion University of the 
Negev, Beersheba, Israel.}

\begin{abstract}

We look at the influence of external fields on systems described by generic
free energy functional of the order parameter. The external force may have
arbitrary spatial dependence, and the order parameter coupling may be
nonlinear. The treatment generalizes seemingly disparate works,
such as pure fluids, liquid and polymer mixtures, lipid
monolayers, and colloidal suspensions in electric fields, fluids, and nematics in
gravity, solutions in an ultracentrifuge, and liquid mixtures in laser radiation.
The phase lines and thermodynamic behavior are calculated at the mean-field
level. We find a ``surface'' critical point that can be shifted to higher or
lower temperatures than the bulk critical point. Below this point, the
transition from a ``gas'' phase to a ``liquid'' phase is first-order, while
above it, the transition is second-order. The second-order line is affected by
the spatial dependence of the force, while the first-order line is universal.
Moreover, the susceptibility may diverge at a finite location ${\bf r}$. Several
analytical expressions are given in the limit where a Landau expansion of the
free energy is valid.

\end{abstract}

\maketitle

\section{Introduction}

The influence of external fields, such as electric \cite{ll_book_elec} or 
magnetic fields \cite{kittel_book}, gravity \cite{moldover_rmp1979}, shear 
\cite{larson_book}, compression and deformation fields 
\cite{ll_elasticity_book}, on given systems has been studied extensively over 
the decades both theoretically and experimentally. A microscopic picture of 
local and global forces enables one to obtain the dynamical and thermodynamic 
behavior on a sufficiently large scale using a coarse-grained order-parameter 
\cite{vandervegt_soft-matter2013,plathe_chemphyschem2002}. While a small field 
may change the system's state slightly, the interesting cases occur when the 
field is large 
enough to induce a phase transition. In most studies, the external field 
is assumed to be (i) constant in space, and (ii) the coupling between field and 
order-parameter is bilinear, e.g., a term $mh$ in classical magnetic (Ising) 
models \cite{huang_book} ($m$ is magnetization and $h$ is the field amplitude), 
or a term $\rho gh$ in gravity ($\rho$ is a fluid's density, $g$ the acceleration on 
Earth, and $h$ a height above an arbitrary reference) \cite{moldover_rmp1979}. 

The influence of Earth's gravity on pure fluids and liquid mixtures has been 
studied thoroughly, and gravity was shown to cause small changes in the behavior 
close to the critical point 
\cite{hohenberg_pra_1972,dickinson_scott_prl1975,greer_prl1975,miller_pra75, 
moldover_rmp1979}. Electric fields have been the subject of equally extensive 
study. Strain due to electrostriction is well documented in solids and fluids 
\cite{moldover_prl1999,moldover_pre1999}. Even in the absence of electrostriction, 
Landau and Lifshitz showed that when a 
simple fluid is subject to a uniform electric field $E$, if the constitutive 
relation between permittivity and density is nonlinear, the critical temperature 
may shift by a small amount, proportional to $E^2~$ \cite{ll_book_elec}. 

The field affects the system's state as it couples to its mass or charge 
densities, magnetic or electrical dipoles, etc. In equilibrium, the system's 
order parameter is given by an equation expressing energy minimization. In 
general, the field itself obeys another equation, and thus, variations in the 
order parameter affect the field self-consistently. In many cases, though, this 
back-action on the field, manifest in the electrostatic case by the existence of 
Laplace or Poisson's equations, can be solved, and the field can be expressed 
explicitly as a function of the order parameter. 

In recent years, several studies have stressed the importance of spatially 
nonuniform fields on the state of simple fluids 
\cite{ttl_nature2004,tsori_rmp2009,vink_archer_pre2012}. When an initially 
homogeneous system is subject to a smoothly varying field, the density, 
composition, magnetization, etc., changes smoothly with coordinate. If the fluid 
is inherently bistable, then a sufficiently strong external field can lead to 
phase separation \cite{tsori_jpcb2011}, nucleation \cite{tsori_jcp2021}, 
wetting-like behavior \cite{tsori_pnas2007,tsori_pre2021}, or filling 
transitions \cite{tsori_jcis-comm2016}. Similar behavior has been reported in 
colloidal suspensions \cite{chaikin_prl2006} and lipid monolayers 
\cite{kycl_science1994,mcconnel_pnas1984} in electric fields, in mixtures in 
laser radiation 
\cite{ducasse_physicaa1999,qui_mm2000,qui_nature_mat2004,wynne_nature_chem2018}, 
in nematic rods in gravity \cite{khokhlov_nematics_pre1999}, in pumped magnetic 
systems \cite{kafri_prl2022}, and in molecular solutions in an ultracentrifuge 
\cite{scriven_jcis1986}.

Here we investigate the thermodynamics of systems exposed to a spatially 
nonuniform external field. The system is described by a generic free energy 
functional depending on the 
coarse-grained order-parameter $\phi$. The expression for the field is quite 
general: the spatial dependence can be arbitrary and the coupling to the 
order-parameter is nonlinear. The current framework thus unifies  
previous works on seemingly disparate systems.

\section{Model}

The starting point of our treatment is the following mean-field free energy 
density, 
\bea\label{eq_bulk}
f=f_b(\phi,T)+\frac12 c^2(\nabla\phi)^2+H(\phi,r)-\mu_0\phi, 
\eea
where $f_b$ is a bulk energy depending on the order-parameter $\phi$ and on 
temperature $T$; it has specific forms for liquid-liquid, liquid-vapor, 
magnetic, or other systems, but we leave it here in a general form. The 
gradient-squared term is essential for describing interfacial and wetting 
phenomena \cite{rolley_rmp2009,nakanishi_fisher_prl1982,leger_joanny_rpp1992, 
andelman_joanny_epl1988}, with $c$ being proportional to a characteristic 
molecular scale. $\mu_0$ is the chemical potential. We are interested in looking 
at the shift to the phase lines at all values of $T$ and especially close to the 
bulk critical temperature $T_c$. $f_b(\phi,T)$ is assumed to be convex at all 
values of $\phi$ when $T>T_c$ and concave at a certain ``window'' of $\phi$ when 
$T<T_c$. $H(\phi,r)$ is a function that contains the influence of the external 
field. The coordinate $r$ is a dimensionless distance in effective 
one-dimensional systems, i.e., in systems with Cartesian, cylindrical, or 
spherical symmetries. There is a location where the external force attains its 
maximum. For the monotonic forces considered below, this point can be thought of 
as being a ``surface'' even in the absence of a physical surface. 

The bulk critical point $(\phi_c,T_c)$ is classically obtained as a solution to 
the two coupled equations: 
$\partial^2f_b(\phi_c,T_c)/\partial\phi^2=\partial^3f_b(\phi_c,T_c)/\partial\phi 
^3=0$ for the two variables $\phi_c$ and $T_c$. A nonuniform force leads to a 
nonuniform profile $\phi(r)$. One may then ask: {\it what is the bulk point 
$(\phi_0,T)$ for which the composition and temperature at the ``surface'' obey 
the critical-point conditions?} 

We focus on external fields that can be written in the form
\bea
H(\phi,r)=h(\phi)q(r).
\eea
For simplicity in the analysis, we assume that $h(\phi)$ is a monotonically 
decreasing or increasing function of $\phi$ and is independent of $T$. In 
addition, $q(r)$ monotonically decreases, e.g., $q(r)=r^{-n}$ 
with $n>0$. The dimensionless distance $r$ is in the range $1\leq r\leq\infty$ 
($r=1$ is the ``surface''). Without loss of generality $q(r=1)=1$ and $0\leq 
q(r)\leq 1$.

\section{Three equations for surface critical point and phase diagram}

The equilibrium profile $\phi(r)$ obeys the Euler--Lagrange 
equation
\bea\label{eq_el}
\frac{\delta f}{\delta\phi}=\frac{\partial 
f_b}{\partial\phi}-c^2\nabla^2\phi+\frac{\partial 
h(\phi)}{\partial\phi}q(r)-\mu_0=0.
\eea
At large distances, $q(r\to\infty)= 0$, the external force vanishes 
and we retrieve a bulk system at composition $\phi_0$; the bulk chemical 
potential is $\mu_0=f_b'(\phi_0,T)$. 
In contrast to classical wetting phenomena 
\cite{cahn_jcp1977,leger_joanny_rpp1992,rolley_rmp2009}, 
Eq. (\ref{eq_el}) cannot be integrated because of the $r$-dependence.
It is instructive to consider cases where $q(r)$ is characterized by a 
spatial scale much larger than the typical interfacial length $\sim c$. In this 
sharp-kink approximation, the gradient-squared term in $f$ can be neglected and 
the free energy is governed by bulk effects. We denote by $\phi_s$ the 
surface composition, i.e. $\phi_s\equiv\phi(r=1)$. When $\phi_s$ satisfies 
the critical-point conditions it obeys the following three equations:
\bea
&&f_b'(\phi_s,T)+h'(\phi_s)q(r)-\mu_0=0,\label{3eqs_eq1}\\
&&f_b''(\phi_s,T)+h''(\phi_s)q(r)=0,\label{3eqs_eq2}\\
&&f_b'''(\phi_s,T)+h'''(\phi_s)q(r)=0.\label{3eqs_eq3}
\eea

We now solve these equations with $q(r)=1$
as follows: We solve Eqs. (\ref{3eqs_eq1}) and 
(\ref{3eqs_eq2}) simultaneously: Eq. (\ref{3eqs_eq2}) gives $\phi_s$; when 
substituted in Eq. (\ref{3eqs_eq1}), the bulk value $\phi_0$ for the given 
temperature can be found from the definition of $\mu_0$.
Similarly, we solve Eqs. (\ref{3eqs_eq1}) and (\ref{3eqs_eq3}) 
simultaneously: $\phi_s$ that solves Eq. (\ref{3eqs_eq3}) can be used in Eq. 
(\ref{3eqs_eq1}) to give the corresponding bulk value $\phi_0$.

\subsection{Solutions of Eqs. (\ref{3eqs_eq1})-(\ref{3eqs_eq3})} 
\begin{figure}[!th]
\begin{center}
\includegraphics[width=0.45\textwidth,clip]{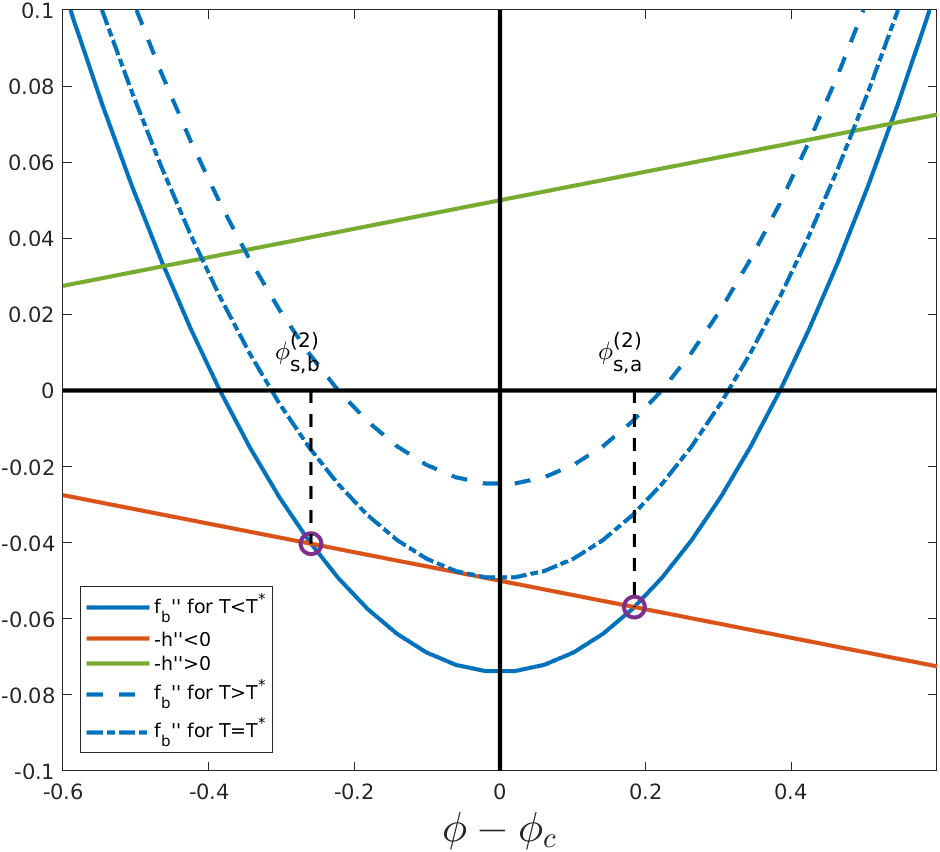}
\caption{Graphical solution of Eq. (\ref{3eqs_eq2}). When $r=1$ this equation 
can be written as $f_b''=-h''$. When the temperature is 
small enough, the intersection of the blue curve ($f_b''$) with the red curve 
($-h''$, assumed negative) gives two solutions, $\phi_{s,a}^{(2)}$ and 
$\phi_{s,b}^{(2)}$, marked by circles. If the temperature is too high, $T_c-T$ 
is small, and there is no solution to Eq. (\ref{3eqs_eq2}) (dashed line). At 
$T=T^*$ (dashed-dot line), $\phi_{s,a}^{(2)}=\phi_{s,b}^{(2)}$ and there is one 
solution to the equation. The green curve shows $-h''$ when it is positive. }
\label{fig_graph_sol_f2p}
\end{center}
\end{figure}

Direct solution of equations (\ref{3eqs_eq1})-(\ref{3eqs_eq3}) is straightforward; 
however, to emphasize the general aspects of the solution and to increase the physical 
insight, we present here a graphical construction. 

{\bf Graphical solution of Eq. (\ref{3eqs_eq2}) and definition of $T^*$.} Figure 
\ref{fig_graph_sol_f2p} shows $f''_b(\phi)$ when $T$ is sufficiently small 
(large $T_c-T$, solid blue line) and $-h''(\phi)$ (red line). The intersection 
of the two curves is the solution to Eq. (\ref{3eqs_eq2}) when $r=1$.  The 
curves intersect at two compositions, denoted $\phi_{s,a}^{(2)}$ and 
$\phi_{s,b}^{(2)}$. We define $\phi_{s,a}^{(2)}> \phi_{s,b}^{(2)}$. The 
subscript a or b denotes either solution; the superscript $2$ signifies that 
they are solutions of the second derivative of $f$. Each of the two solutions 
$\phi_{s,i}^{(2)}$ (i = a, b) has a different bulk value of $\phi_0$, obtained when 
it is substituted in Eq. (\ref{3eqs_eq1}). We call these 
values $\phi_{0,a}^{(2)}$ and $\phi_{0,b}^{(2)}$.

A different behavior occurs if the temperature is not low enough (alternatively, 
if the strength of the external force $|h''|$, is too large), as indicated by 
the dashed blue line -- in this case, the curves do not intersect, hence there is 
no solution to Eq. (\ref{3eqs_eq2}). We now define $T^*$ as the 
temperature at which $\phi_{s,a}^{(2)}$ and $\phi_{s,b}^{(2)}$ are equal, see 
dashed-dot line in Fig. \ref{fig_graph_sol_f2p}.
\begin{figure}[!th]
\begin{center}
\includegraphics[width=0.45\textwidth,clip]{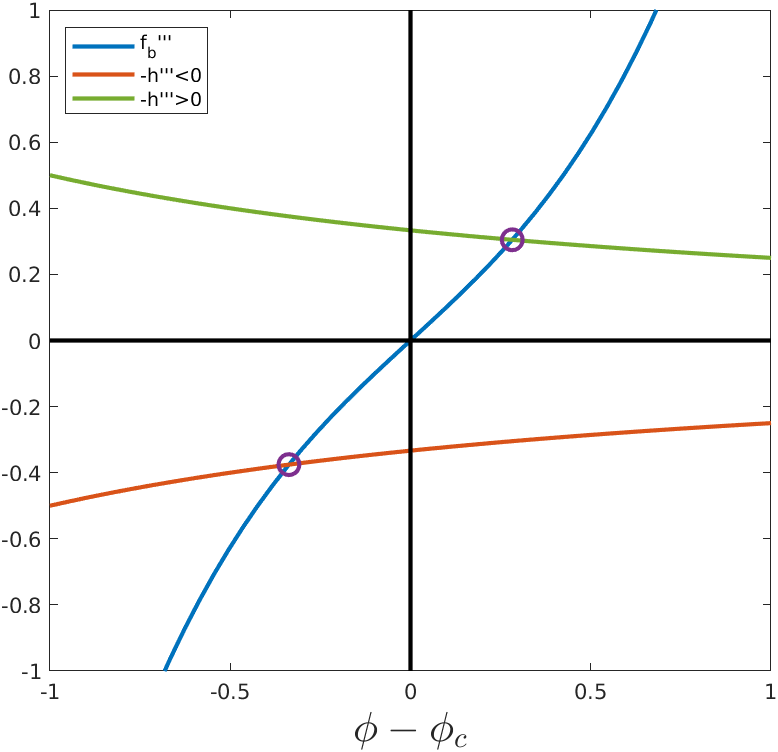}
\caption{Graphical solution of Eq. (\ref{3eqs_eq3}). When $r=1$ this equation 
can be written as $f_b'''=-h'''$. The $f_b'''$ curve 
looks typically like the blue curve. Two curves are shown for $h'''$ -- for 
the two cases where $h'''$ is either positive or negative. In both 
cases, there is always a solution (circles).}
\label{fig_graph_sol_f3p}
\end{center}
\end{figure}
\begin{figure}[!th]
\begin{center}
\includegraphics[width=0.45\textwidth,clip]{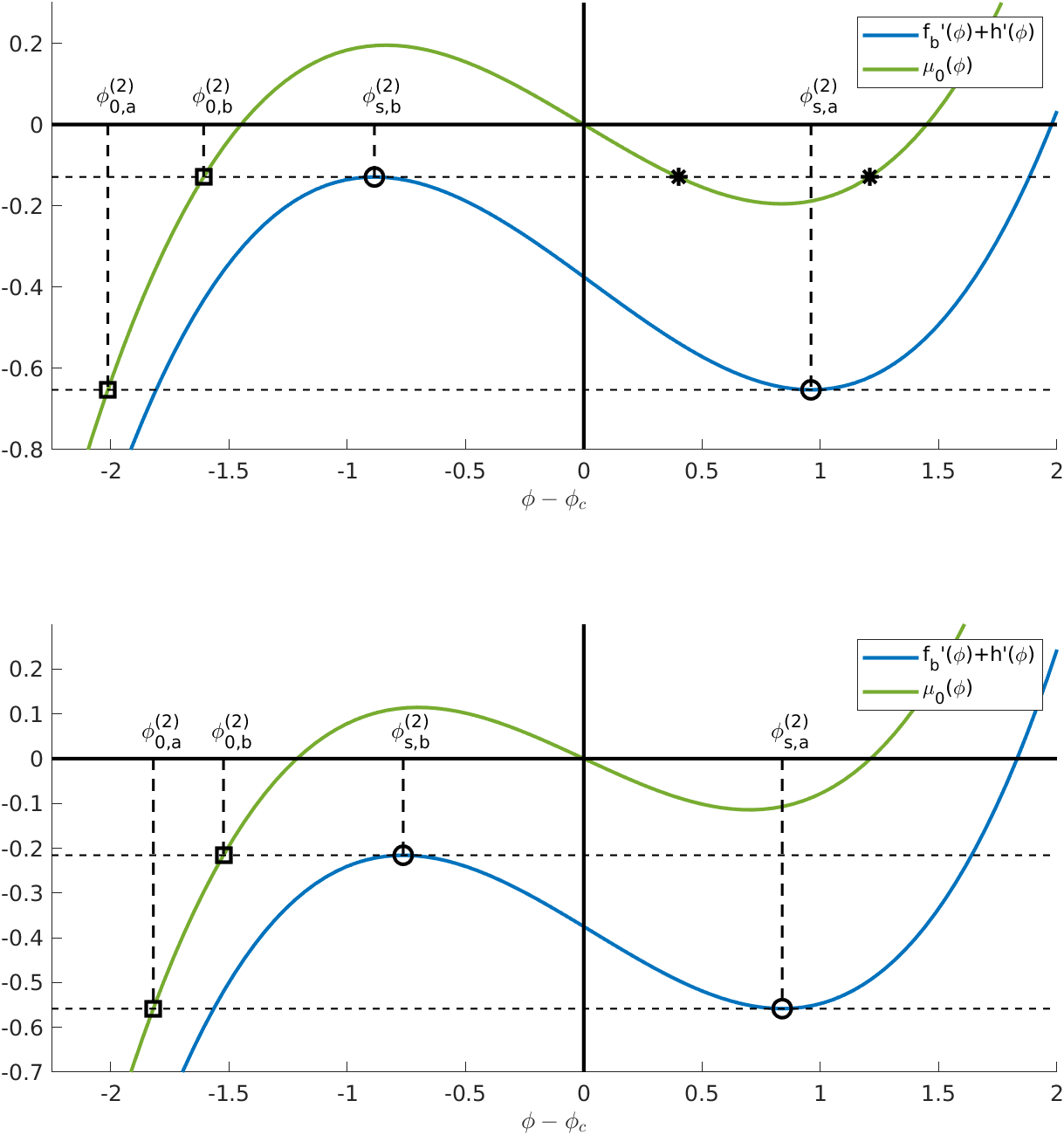}
\caption{Graphical solution of Eq. (\ref{3eqs_eq1}). Blue curve is 
$f_b'(\phi)+h'(\phi)$ and the green curve is $\mu_0(\phi)$. The circles mark the 
solutions of Eq. (\ref{3eqs_eq2}). The values of $\phi_{0,a}^{(2)}$ and 
$\phi_{0,b}^{(2)}$ can be obtained as the intersection of the horizontal lines 
with the $\mu_0(\phi)$ curve. The top and bottom panels are for low 
and high temperatures, respectively (both lower than $T^*$). 
}
\label{fig_finding_phi0}
\end{center}
\end{figure}

{\bf Graphical solution of Eq. (\ref{3eqs_eq3}).}
The graphical solution is shown in Fig. \ref{fig_graph_sol_f3p}. Here, it is easy to see 
that one solution $\phi_s^{(3)}$ always exists. Once this root is known, it can be 
substituted in Eq. (\ref{3eqs_eq1}) to yield the corresponding bulk value 
$\phi_0^{(3)}$.  
\begin{figure}[!th]
\begin{center}
\includegraphics[width=0.45\textwidth,clip]{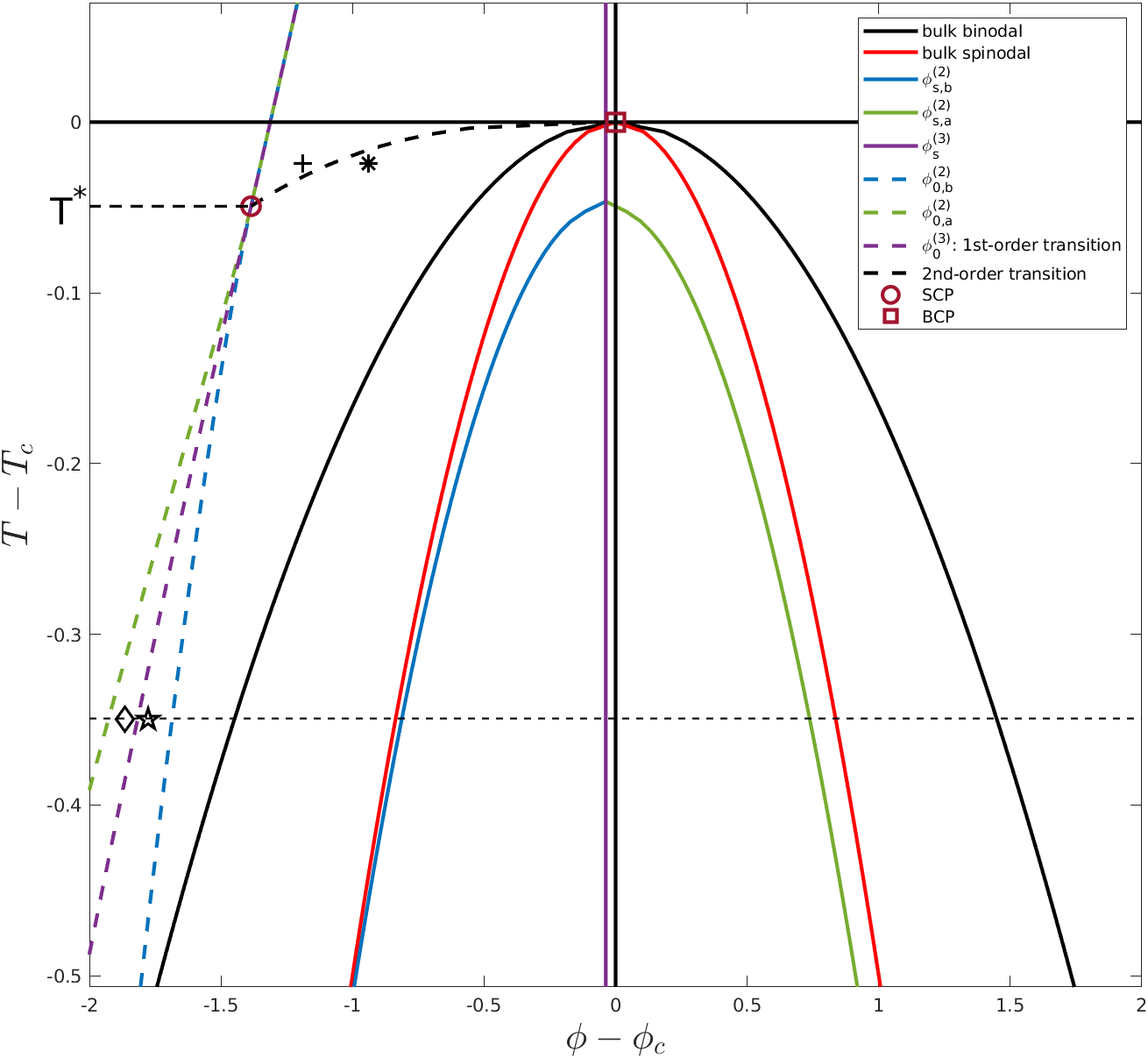}
\caption{Phase diagram in the $\phi-T$ plane when $-h''$ is negative (as the 
red curve in Fig. \ref{fig_graph_sol_f2p}). The bulk critical point (BCP) is at 
$(\phi_c,T_c)$ (square). The bulk binodal and spinodal curves are in black and 
red, respectively. The curves $\phi_{s,a}^{(2)}(T)$, $\phi_{s,b}^{(2)}(T)$, and 
$\phi_{s}^{(3)}(T)$ are in solid green, blue and purple, while 
$\phi_{0,a}^{(2)}(T)$, $\phi_{0,b}^{(2)}(T)$, and $\phi_{0}^{(3)}(T)$ are in 
identical colors and dashed lines. 
$\phi_{0}^{(3)}(T)$ is a first-order transition line (dashed purple curve). The 
surface critical point (SCP), marked with a circle, is at a temperature below 
the critical point ($T^*<T_c$). The second-order line is the dashed thick black 
curve. The phase diagram was calculated using the Landau expansion of $f_b$ in Eq. 
(\ref{landau_expansion}) with $d=1$ and the wedge external field in Eqs. 
(\ref{wedge_fe}) and (\ref{epsilon_expansion}), with $\veps_c=7$, $
\veps_1=3$, $\veps_2=-0.4$, $\veps_3=-0.3$, and $M=0.25$. 
}
\label{fig_phase_diagram1}
\end{center}
\end{figure}

{\bf Graphical solution of Eq. (\ref{3eqs_eq1}).} Figure \ref{fig_finding_phi0} 
shows the graphical construction for finding the value of $\phi_{0,a}^{(2)}$ and 
$\phi_{0,b}^{(2)}$. The figure has two parts, for two different temperatures -- 
top panel is for low $T$ and the bottom is for higher $T$ (both smaller than $T^*$).
There are two curves in each part,
representing two parts of Eq. (\ref{3eqs_eq1}). The blue 
curve is $f_b'(\phi)+h'(\phi)$ and the green curve is $\mu_0(\phi)$. The two 
circles mark the values of $\phi_{s,a}^{(2)}$ and $\phi_{s,b}^{(2)}$. 
The solution to Eq. (\ref{3eqs_eq1}) is 
obtained as the intersection of the horizontal line at each of these points with 
the green curve. At the low temperature, there are three possible values of 
$\phi_{0,b}^{(2)}$. The correct solutions are marked with squares, the two 
other solutions are marked with ``*''. At the higher temperature (bottom 
panel), there is only one possible value of $\phi_{0,b}^{(2)}$.

{\bf Phase diagram.} The resulting phase diagram is shown in Fig. 
\ref{fig_phase_diagram1}. For sufficiently low temperatures, two solutions to 
Eq. (\ref{3eqs_eq2}) exist, $\phi_{s,a}^{(2)}(T)$ and $\phi_{s,b}^{(2)}(T)$ 
(green and blue, respectively), and usually they are below the spinodal curve. 
As the temperature is elevated, the two solutions approach each other. At  
$T=T^*$, they merge. The corresponding bulk 
composition curves $\phi_{0,a}^{(2)}(T)$ and $\phi_{0,b}^{(2)}(T)$ (dashed green 
and blue) behave similarly, and merge at $T^*$ at a ``surface critical point'' 
(SCP), marked by a large circle. At this point, the bulk values $\phi_0$ and $T$ 
are such that, {\it at the surface}, the second and third derivative of the 
energy with respect to $\phi$ both vanish.

The curve $\phi_0^{(3)}$ is a first-order phase line; crossing it from low to 
high values of $\phi_0$ at constant $T$ leads to a discontinuous change in the 
profile $\phi(r)$. This line is universal and does not depend on the exact form 
of the spatial dependence of the external force, $q(r)$.

\begin{figure}[!th]
\begin{center}
\includegraphics[width=0.45\textwidth,clip]{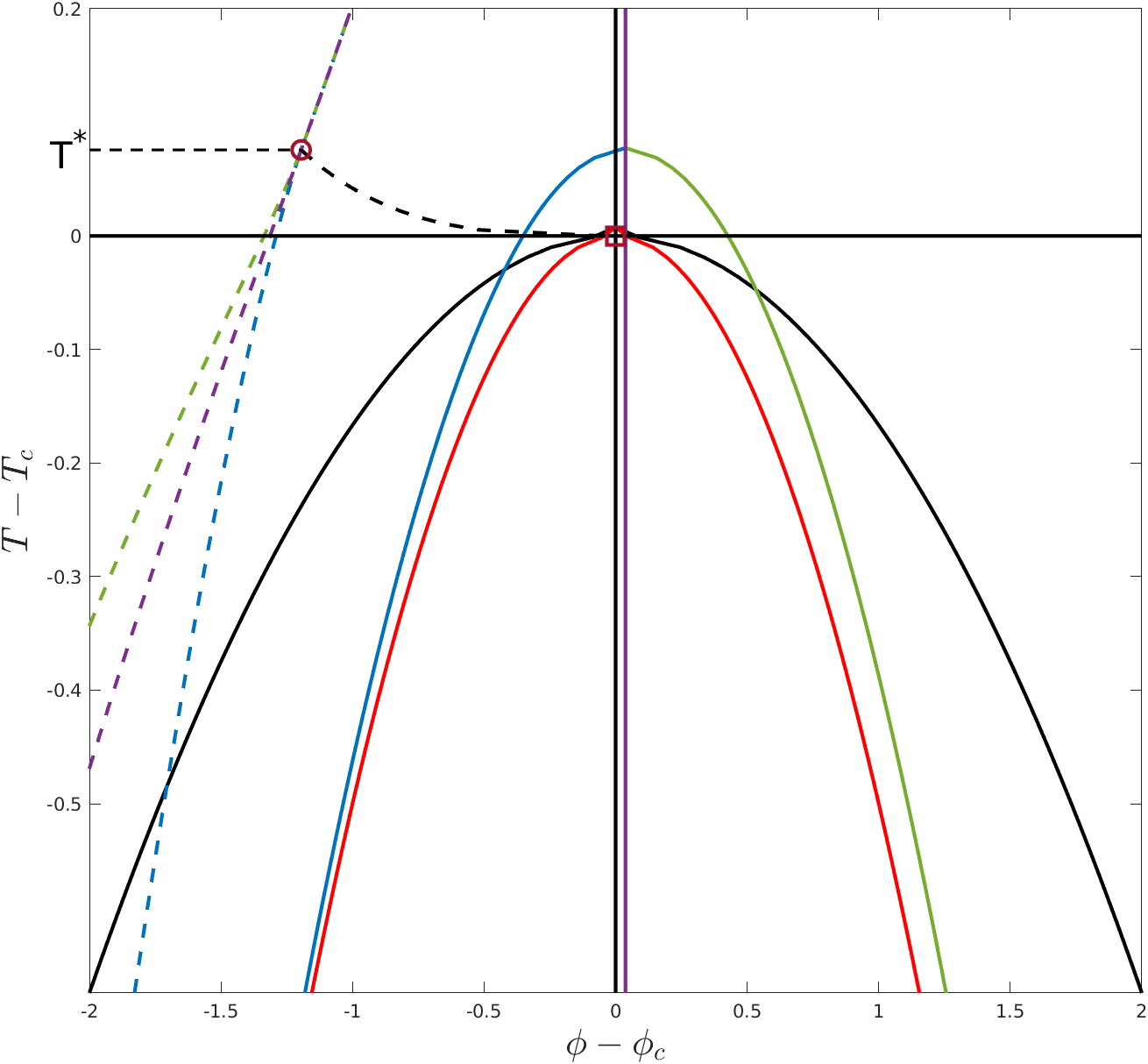}
\caption{The same as in Fig. \ref{fig_phase_diagram1} but this time $-h''$ is 
positive in the relevant range (green curve in Fig. \ref{fig_graph_sol_f2p}). The 
surface critical point is displaced upwards and $T^*>T_c$. The figure was calculated 
using the same model and parameter values as in Fig. \ref{fig_phase_diagram1},  
except $\veps_2=0.6$ and $\veps_3=0.3$.
}
\label{fig_phase_diagram2}
\end{center}
\end{figure}

The SCP point depends on the amplitude of the external forcing. The point 
displaces when the amplitude of the external force reduces toward zero. For any 
finite value of $r$, Eqs. (\ref{3eqs_eq1})-(\ref{3eqs_eq3}) yield the triplet 
solution $[\phi_s(r),\phi_0(r),T(r)]$. The second-order phase transition line, 
dashed black curve, is the locus of points $[\phi_0(r),T(r)]$ for $r$ increasing 
from $r=1$ (SCP) to $r=\infty$ (BCP). Crossing it leads to a continuous change 
in the profile $\phi(r)$. On the second-order line, Eq. (\ref{3eqs_eq2}) is 
satisfied at a finite distance from it, at $r=r_0>1$ that depends on 
temperature. Fig. \ref{fig_phase_diagram2} is similar but here $-h''$ is positive and 
the SCP is found at temperatures higher than $T_c$. 

One can now see that the SCP point and the phase transition lines would not 
appreciably change even if the $(\nabla\phi)^2$ term in the energy were 
included. Indeed, if Eq. (\ref{3eqs_eq1}) has a $\phi''$ 
term, just before the transition, the profiles are smoothly varying with $r$ 
and this second derivative is small. In addition, by definition, the interface 
separating high and low values of $\phi$ is an inflection point in the profile so 
$\phi''=0$.

\begin{figure}[!th]
\begin{center}
\includegraphics[width=0.45\textwidth,clip]{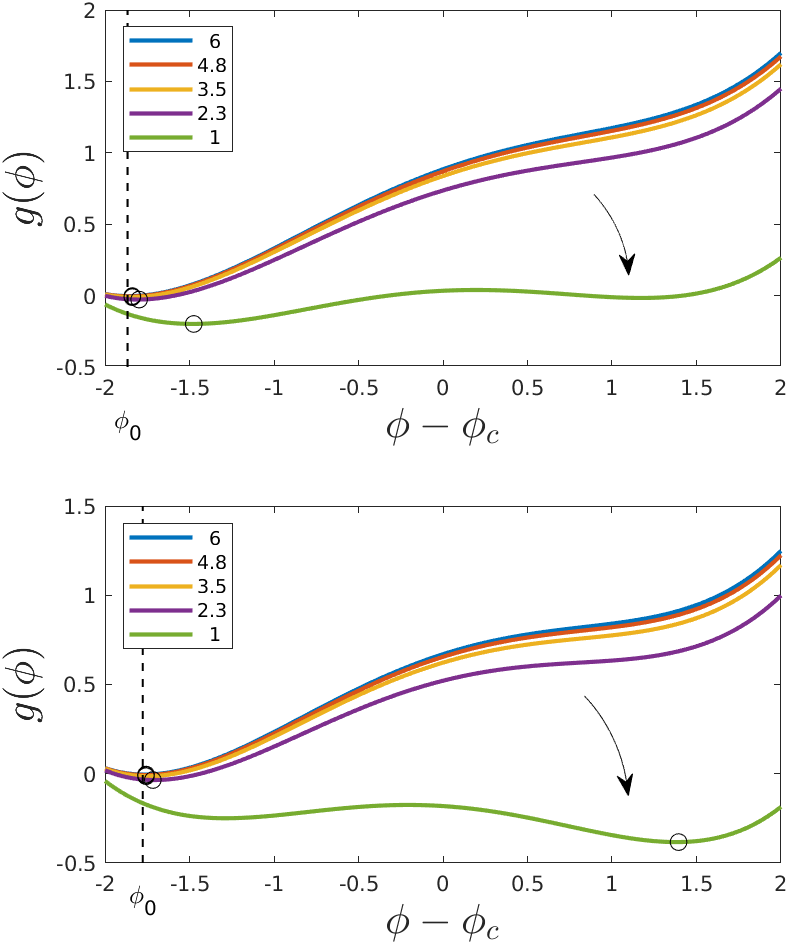}
\caption{Plots of the function $g(\phi)$ given in Eq. (\ref{g_of_phi}) for 
several values of $r$. Top: when the external force is small, the bulk 
composition $\phi_0$ is outside of the first-order transition line (e.g. point 
``$\Diamond$'' in Fig. \ref{fig_phase_diagram1}). As $r$ decreases from $\infty$ to 
$1$ ($r$ values are in the legend), the minimum of $g(\phi)$, indicated by circles, 
shifts to larger values. However, the minimum stays close to the bulk value, $\phi_0$. 
The resulting profile $\phi(r)$ is shown in Fig. \ref{fig_profiles_below_Tstar} (top). 
Bottom: when the external forcing is large, $\phi_0$ is inside the stability line (e.g. 
point ``$\bullet$'' in Fig. \ref{fig_phase_diagram1}). As $r$ decreases, the minimum 
of $g(\phi)$ jumps from small to large values of $\phi$. See the profile in Fig. 
\ref{fig_profiles_below_Tstar} (bottom).
}
\label{fig_g_of_phi}
\end{center}
\end{figure}

\section{Construction of profiles}
\begin{figure}[!th]
\begin{center}
\includegraphics[width=0.45\textwidth,clip]{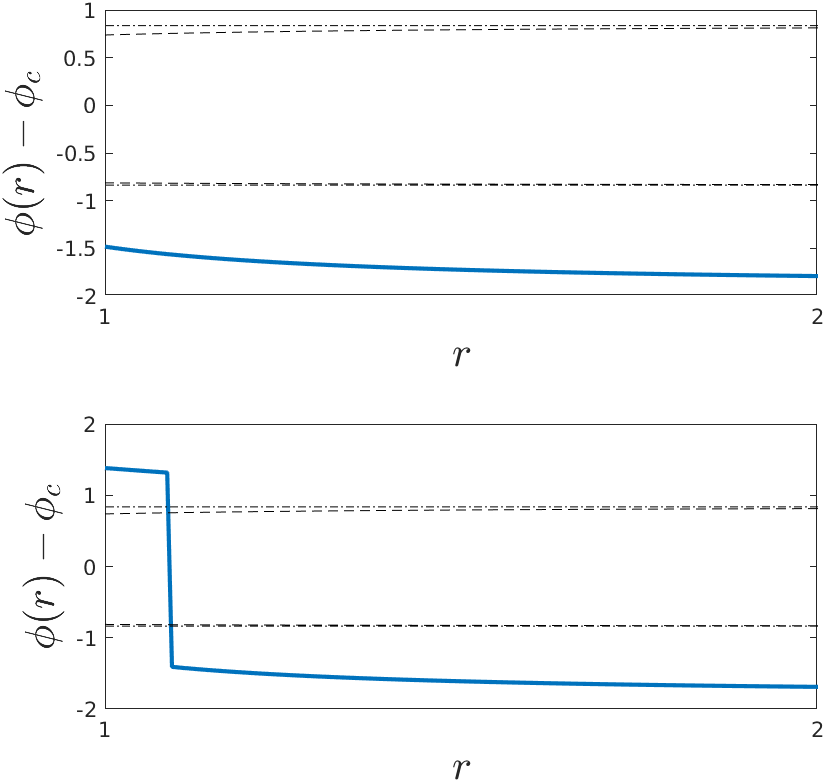}
\caption{Composition profiles at a temperature below $T^*$ in the 
sharp kink approximation. The top and bottom 
panels correspond to bulk points marked by ``$\Diamond$'' and ``$\bullet$'', 
respectively, in Fig. \ref{fig_phase_diagram1}. While at point ``$\Diamond$'' 
$\phi(r)$  is smoothly varying, at ``$\bullet$'' it exhibits a discontinuity. The 
transition from the top curve to the bottom one is discontinuous (first-order).
At any given value of $r$, there are two values of $\phi$ that obey 
$f''(\phi,r)=0$ [Eq. (\ref{eq_f2p_vs_r})]. The dashed lines show these values 
vs. $r$. The horizontal dash-dot lines are the bulk spinodal values at this 
temperature (i.e. solutions of $f_b''(\phi,T)=0$).
}
\label{fig_profiles_below_Tstar}
\end{center}
\end{figure}

To construct the composition profiles, we look at the function $g$ defined by
\bea\label{g_of_phi}
g(\phi)\equiv f_b(\phi)+h(\phi)q(r)-\mu_0\phi-p_0,
\eea
where $p_0=f_b(\phi_0)-\mu_0\phi_0$ is the bulk pressure. The ``correct'' value 
of composition at a specific $r$ is the value of $\phi$ that minimizes 
$g(\phi)$. Next, we concentrate on two points, marked in Fig. 
\ref{fig_phase_diagram1}: the bullet is ``inside'' the first-order line, and the 
diamond is outside. The grand-potential $g(\phi)$ plots for the two points are 
shown in Fig. \ref{fig_g_of_phi} for several values of $r$. The top panel shows 
the $g(\phi)$ curves when the bulk composition is outside of the first-order 
line (diamond). When $r$ is large, the minimum is at $\phi_0$ (dashed vertical 
line). As $r$ decreases, the minimum gradually shifts to larger values of 
$\phi$. Yet, even at the smallest possible value of $r$ ($r=1$) the minimum 
stays rather close to $\phi_0$. The behavior is different at the bullet point, 
found below the first-order line. The bottom panel shows the $g(\phi)$ curves in 
this case. Here, as $r$ decreases from large to small values, a discontinuous 
transition occurs between low and high values of $\phi$.
\begin{figure}[!th]
\begin{center}
\includegraphics[width=0.45\textwidth,clip]{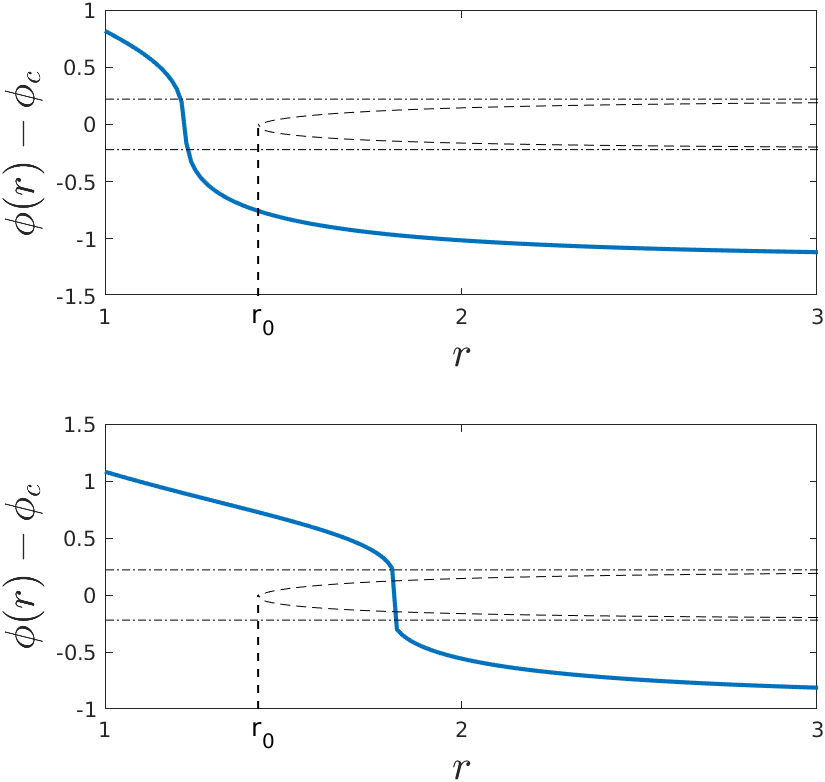}
\caption{Composition profiles at a temperature above $T^*$. Top and bottom 
panels correspond to bulk points marked by ``+'' and ``$\times$'', respectively, in 
Fig. \ref{fig_phase_diagram1}. While at the point ``+'', $\phi(r)$  
is smoothly varying, at ``$\times$'' it exhibits a discontinuity. However, the transition 
from the top curve to the bottom one is continuous (second-order).
At any given value of $r$, there are two values of $\phi$ that obey 
$f''(\phi,r)=0$ [Eq. (\ref{eq_f2p_vs_r})]. The dashed lines show these 
values vs $r$. The two curves merge at $r=r_0$. The horizontal dash-dot lines 
are the bulk spinodal values at this temperature (i.e. solutions of 
$f_b''(\phi,T)=0$).
}
\label{fig_profiles_above_Tstar}
\end{center}
\end{figure}
The corresponding profiles $\phi(r)$ are shown in Fig. 
\ref{fig_profiles_below_Tstar}. 

We now look at two points at a temperature above $T^*$, marked by ``+'' and 
``$\times$'' in Fig. \ref{fig_phase_diagram1}. Across the dashed black line in the 
figure, the transition is second-order. Indeed, Fig. \ref{fig_profiles_above_Tstar} 
shows the corresponding $\phi(r)$ curves. When the point is to the left (``+''), the 
profile is continuous (though it might have a relatively steep domain). When the point 
is to the right (``$\times$''), the profile becomes discontinuous. The difference 
between this scenario and the previous one is that here the discontinuous profile 
emerges from the continuous one as a second-order phase transition. The 
second-order line is shown as a dashed black curve in Fig. \ref{fig_phase_diagram1}. 

In Figs. \ref{fig_profiles_below_Tstar} and \ref{fig_profiles_above_Tstar}, we 
show two dashed black lines. They correspond to the two possible solutions of 
$f''(\phi)=0$, that is, they are solutions of 
\bea
f_b''(\phi)+h''(\phi)q(r)=0\label{eq_f2p_vs_r}
\eea
for a specific value of $r$. For low temperatures, $T<T^*$, the two curves only 
meet at some $r$ smaller than unity, which is not valid. For high temperatures, 
$T>T^*$, they meet at a finite value of $r>1$, denoted $r_0$. When the jump in 
the profile occurs exactly at $r=r_0$, one is on the second-order phase 
transition line. In Fig. \ref{fig_profiles_above_Tstar} (bottom) the jump is for 
$r>r_0$ and therefore in this example we are already past the phase transition 
line (point ``$\times$''). For a given $T$ and amplitude of external force, the 
second-order line can thus be obtained when $\phi_s$ is the single solution to 
Eq. (\ref{eq_f2p_vs_r}), with $r$ varying from $1$ to $\infty$, and $\phi_0$ 
obtained from Eq. (\ref{3eqs_eq1}). The SCP is retrieved as the special case 
when $r_0=1$. 

{\bf The susceptibility $\chi$}. There are several ways to define the 
susceptibility in spatially-varying fields. We rewrite the sharp-kink profile 
equation as
\bea
f_b'(\phi)+h'(\phi)q(r)-\mu_0=0.
\eea
and interpret $-h'(\phi)q(r)$ as the field. A derivative of the profile 
equation with respect to the field then gives
\bea\label{eq_chi}
\chi=\frac{1}{f_b''(\phi(r))}.
\eea
Importantly, $\chi$ diverges at a point $r$ if $\phi(r)$ equals one of the bulk 
spinodal values.
\begin{figure}[!th]
\begin{center}
\includegraphics[width=0.45\textwidth,clip]{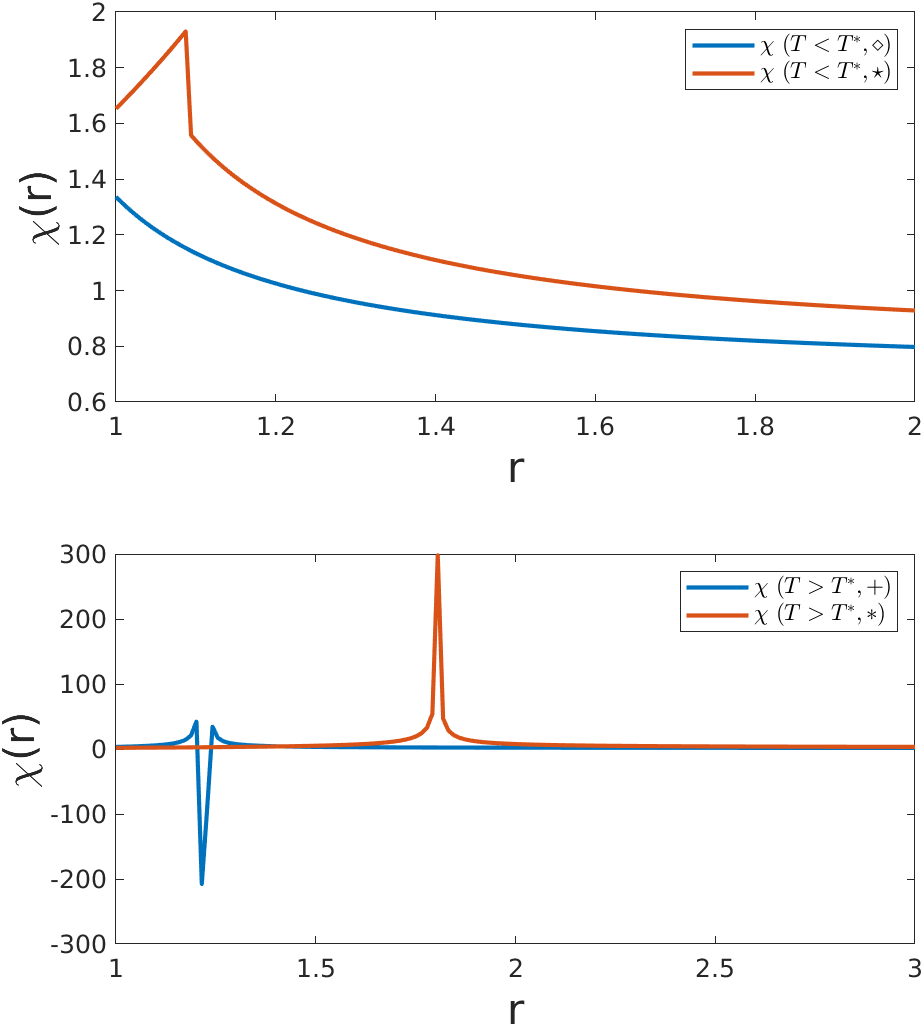}
\caption{Susceptibility $\chi$ defined by Eq. (\ref{eq_chi}) and applied to the 
profiles $\phi(r)$ from Figs. \ref{fig_profiles_below_Tstar} and 
\ref{fig_profiles_above_Tstar}. $\chi$ stays finite for the profile at point 
``$\diamond$'' in the phase-diagram Fig. \ref{fig_phase_diagram1} and diverges 
at the points ``$\bullet$'', ``+'', and ``$\times$''. 
}
\label{fig_suscept}
\end{center}
\end{figure}
Figure \ref{fig_suscept} shows $\chi$ calculated by Eq. (\ref{eq_chi}) for the 
profiles shown in Figs. \ref{fig_profiles_below_Tstar} and 
\ref{fig_profiles_above_Tstar}. $\chi$ diverges for the profiles $\phi(r)$ that 
cross one of the two spinodal values (points ``$\bullet$'', ``+'', and ``$\times$'') and 
remains finite otherwise (point ``$\diamond$'').

Note that the susceptibility can tend to $+\infty$ at one side of a point $r$ 
and to $-\infty$ at the other side. This happens when the curve $\phi(r)$ 
crosses the bulk spinodal value. The negative susceptibility appears in a small 
region of space [of order $\sim c$ of Eq. (\ref{eq_bulk})] where the 
composition crosses between two phases stabilized by the field. It is 
indicative that the local value of $\phi$ is unstable as a bulk phase.

\section{Landau expansion of the free energy}\label{section_landau}

To get explicit expressions for the quantities 
presented in the previous sections valid close enough to the bulk critical 
point, we use a standard (symmetric) Landau expansion of the free energy and 
external field,
\bea\label{landau_expansion}
f_b&=&f_c+\frac12 t\vphi^2+\frac{1}{24}d\vphi^4,\nn\\
h&=&h_c+h_1\vphi+\frac12 h_2\vphi^2+\frac16 h_3\vphi^3, 
\eea
with $t\sim (T-T_c)/T_c$, and $d$ a positive constant independent of $T$. 
Quantities are 
further expressed in terms of $\vphi=\phi-\phi_c$ instead of $\phi$. 
The bulk chemical potential is 
\bea
\mu_0=t\vphi_0+\frac{1}{6}d\vphi_0^3.\nn
\eea
Equations (\ref{3eqs_eq1})-(\ref{3eqs_eq3}) can now be written for the surface 
[$q(r)=1$] as
\bea
t\vphi_s+\frac16d\vphi_s^3+h_1+h_2\vphi_s+\frac12h_3\vphi_s^2-\mu_0=0,
\label{3eqs_eq1_landau}\\
t+\frac12d\vphi_s^2+h_2+h_3\vphi_s=0,\label{3eqs_eq2_landau}\\
d\vphi_s+h_3=0.\label{3eqs_eq3_landau}
\eea
The solutions to these equations give the qualitative phase diagram in Fig. 
\ref{fig_phase_diagram1} if $h''=h_2+h_3\vphi>0$ and 
$h'''=h_3>0$ or Fig. \ref{fig_phase_diagram2} if 
$h_2+h_3\vphi<0$ and $h_3<0$.

To find the SCP point, we use 
$\vphi_{0,a}^{(2)}=\vphi_{0,b}^{(2)}=\vphi_{0}^{(3)}$ at the SCP, 
and at the SCP temperature, 
$t^*$, $\vphi_{s,a}^{(2)}=\vphi_{s,b}^{(2)}=\vphi_{s}^{(3)}$,
so this composition can be simply called $\vphi_s$. From Eq. 
(\ref{3eqs_eq3_landau}) we find that 
\bea
\vphi_s=-h_3/d. 
\eea
$\vphi_s$ is independent of temperature. 
Hence the line $\vphi_s^{(3)}$ is entirely to the left of the critical composition
if $h_3>0$ and to the right otherwise. Substitution in Eq. (\ref{3eqs_eq2_landau}) 
gives for $t=t^*$,
\bea
t^*&=&\frac12\frac{h_3^2}{d}-h_2.\label{tstar_landau}
\eea
Interestingly, $t^*$ depends on the amplitude of the external forcing 
(expressed via $h_2$ and $h_3$) both linearly and quadratically. $T^*$ is above 
the critical point temperature if 
\bea
\frac12\frac{h_3^2}{d}-h_2>0\nn
\eea
and below otherwise. 

The first-order line $\vphi_0^{(3)}(t)$ is obtained from Eq. 
(\ref{3eqs_eq1_landau}) as the solution of the third-order polynomial
in $\varphi_0$,
\bea
t\vphi_s+\frac16 
d\vphi_s^3+h_1+h_2\vphi_s+\frac12 h_3\vphi_s^2
-t\vphi_0-\frac16 d\vphi_0^{3}=0\nn
\eea
with $\vphi_s=-h_3/d$. 
The composition of the SCP point, $\vphi_0^*$, is a special case with
$t=t^*$, simplifying to 
\bea
h_1-\frac16\frac{h_3^3}{d^2}-t^*\vphi_0^*-\frac16 
d\vphi_0^{*3}=0.
\eea

The lines $\vphi_{s,a}^{(2)}(t)$ and $\vphi_{s,b}^{(2)}(t)$ in the phase 
diagram 
are the solutions to the quadratic equation 
$f''(\vphi,r=1)=0$ [Eq. (\ref{3eqs_eq2_landau})] with varying values of $t$. 
The dashed curves in Fig. \ref{fig_profiles_above_Tstar} are the solution to 
$f''(\vphi_s,r)=0$, namely, 
\bea
t+\frac12d\vphi_s^2+ 
\left(h_2+h_3\vphi_s\right)q(r)=0.\label{eq_fpp_vs_r_landau}
\eea
The value of $r_0$ is defined such that the two solutions $\vphi_s$ are the 
same, i.e., when the discriminant vanishes: $\Delta^2=0$. This yields a 
quadratic equation for $q(r_0)$: 
\bea\label{eq_disc_landau}
\Delta^2=h_3^2q^2(r_0)-2dh_2q(r_0)-2dt=0.
\eea
(if $r=1$ then $t=t^*$). At $t<t^*$, there is no solution with $r_0>1$ 
because $|q(r_0)|<1$, while at $t>t^*$ there is one. The value of 
the single solution $\vphi_s$ is then 
\bea\label{eq_single_sol_landau}
\vphi_s=-h_3q(r_0)/d.
\eea
How to obtain the second-order transition line? This line is obtained by 
increasing $r$ from $1$ to $\infty$ as a parameter, with $t$ given from Eq. 
(\ref{eq_disc_landau}), $\vphi_s$ from Eq. (\ref{eq_single_sol_landau}) (replace 
$r_0\to 
r$), and both substituted in  
\bea
t\vphi_s+\frac16d\vphi_s^3+h_1+h_2\vphi_s+\frac12 h_3\vphi_s^2
-t\vphi_0-\frac16 d\vphi_0^3=0,\nn\\
\eea
to find $\vphi_0$.

%
\begin{figure}[!th]
\begin{center}
\includegraphics[width=0.45\textwidth,clip]{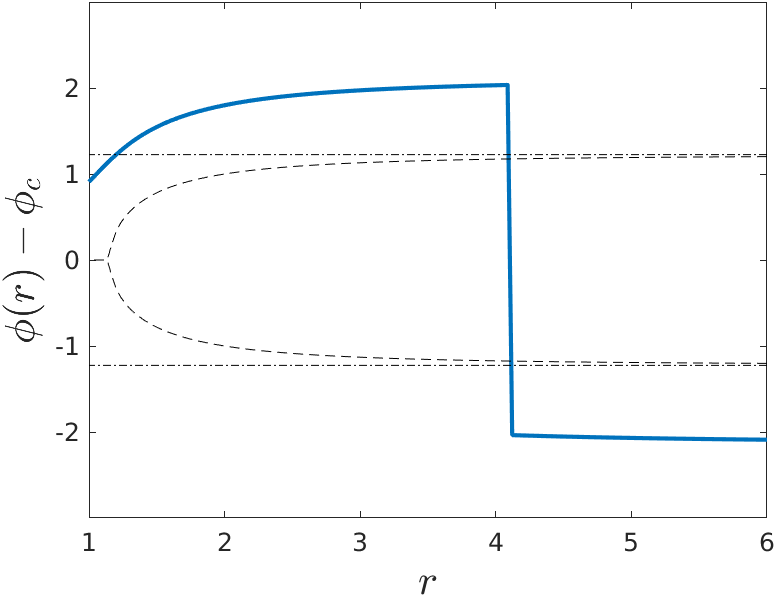}
\caption{Non-monotonic profiles $\phi(r)$ that occur when $h_1<0$ and $h_2>0$ 
is sufficiently large. In the limit $r\to\infty$, the free energy $f(\phi)$ has 
a shape of a double well, with the global minimum at the ``gas'' phase. When 
$r\to 1$, the external field dominate and $f(\phi)$ is approximately
the parabola $h_1\vphi+(1/2)h_2\vphi^2$, whose minimum is at 
$-h_1/h_2$. Hence the decrease in $\phi$ from its value at the liquid-gas 
interface to its value at $r=1$. 
}
\label{fig_nonmonotonic_profile}
\end{center}
\end{figure}

{\bf Non-monotonic profiles.} 
The profiles $\phi(r)$ usually look like the 
ones in Figs. \ref{fig_profiles_below_Tstar} and 
\ref{fig_profiles_above_Tstar}: 
$\phi(r)$ is a monotonically decreasing function, with $\phi(r=\infty)=\phi_0$. 
If $\phi_0$ is within the unstable area of the phase diagram, $\phi(r)$ 
exhibits a sharp drop at the ``liquid''-``gas'' interface. If, however, $h_1<0$ 
and $h_2>0$ are sufficiently large, close to the surface $(r\gtrsim 1$) 
the external field dominates over 
the bulk free energy and $\phi(r)$ can {\it increase} with $r$ inside the 
liquid phase. To see this, assume that $h_1$ and $h_2$ are very large, then at 
$r=1$ the free energy is $f\approx h_1\vphi+(1/2)h_2\vphi^2$, and in this 
parabola, the linear and quadratic terms are negative and positive, 
respectively. 
The minimum value, $-h_1/h_2$, can have any positive value and, in particular, 
it can be smaller than the (upper) spinodal value. Figure 
\ref{fig_nonmonotonic_profile} shows such a profile. 

The free-energy picture as $r$ decreases is this: at $r\to\infty$, 
$f(\phi)$ has a double minima, with the most stable one being the bulk ``gas'', 
at $\phi_0$. As $r$ decreases, the ``liquid'' minimum 
lowers until, at the interface, the two minima are in equal depth. As $r$ 
further decreases, the ``liquid'' minimum becomes the global minimum; it 
continuously shifts with $r$. If $h_2=0$ then it shifts to the right. If $h_2$ 
is sufficiently large, it shifts to the left, while the $f(\phi)$ curve loses 
the ``gas'' minimum and becomes the parabola mentioned above. 

\section{Examples}\label{examples}

We list here several different systems to which the theory presented above applies.

{\bf Centrifugation.} Consider a binary mixture with composition $\phi$ of two 
partially miscible liquids ``1'' and ``2'' placed in a rapidly rotating 
centrifuge. Denoting the angular velocity as $\omega$ and the distance from the 
rotation axis as $R$, we write the effective energy as  
\cite{tsori_crphysique_2007,horinek_pccp_2021}
\bea
f=f_b(\phi)-\frac12 M(\phi) r^{-2}.\nn
\eea
Here, $r=R_{\rm max}/R$, $R_{\rm max}$ is the maximum distance from the 
rotation 
axis, $M=v_0\rho(\phi)\omega^2R_{\rm max}^2/k_BT$,
$v_0$ is a molecular volume, $k_BT$ is the thermal energy, and 
$\rho(\phi)$ is the composition-dependent mass density. Thus 
$h(\phi)=-(1/2)M(\phi)$ and $q(r)=r^{-2}$. Usually, $\rho(\phi)$ is 
approximated as a linear interpolation between the densities of the two 
liquids; in the language of the Landau expansion above, $h_1$ is the 
difference in the liquids' densities, and $h_2=h_3=0$. This means that the 
temperature of the SCP point is $T^*=T_c$ [Eq. (\ref{tstar_landau})]. The SCP 
occurs at the maximal radius of the centrifuge.

{\bf Lipid monolayers.} A monolayer consisting of a mixture of two dipolar 
lipids forms a homogeneous phase if the surface pressure is sufficiently small 
\cite{kycl_science1994,mcconnel_pnas1984}. When a wire passing 
perpendicular to the monolayer is 
charged, smooth density variations appear. However, if the wire's charge (or 
potential) is large, a distinct dense, liquid-like, phase appears around the 
wire. This system's energy is given by 
\bea
f=f_b(\phi)-\phi Mr^{-1},\nn
\eea
where $\phi$ is the lipid density scaled by $v_0^{-1}$, 
$M=p\sigma/\veps k_BT$, $p$ is the molecule's electric dipole moment, 
$\sigma$ is the wire's charge density, $\veps$ is the dielectric constant, 
and $r=R_w/R$ is the ratio between the wire's radius and the distance to the 
wire. We find that $q(r)=r^{-1}$ and $h(\phi)=-\phi M$ is linear in 
$\phi$, leading again to $T^*=T_c$. The SCP occurs at the surface of the wire.

{\bf Wedge capacitor.} 
The wedge capacitor is made up of two semi-infinite metallic planes that have 
an opening angle $\beta$ between them \cite{ttl_nature2004} and the potential 
difference between the plates is $V$. A purely dielectric fluid of 
composition $\phi$ is placed 
between these planes. The electrostatic energy density is  
$-(1/2)\veps E^2$ and the electric field satisfying Laplace's equation is 
$|E|=V/(\beta R)$, hence $h(\phi)$ and $q(r)$ are given by
\bea\label{wedge_fe}
h(\phi)=-\frac12M\veps(\phi)~~~~~~~~q(r)=r^{-2}.
\eea
Here $M=V^2v_0\veps_0/(\beta^2R_{\rm min}^2k_BT)\propto V^2$
denotes the magnitude of the dielectrophoretic force. 
$\veps_0$ is the vacuum permittivity, $r=R/R_{\rm min}$ is 
scaled by $R_{\rm min}$, the minimal value of $R$.

In the vicinity of the critical point, one writes $\veps(\phi)$ as a power 
series around the critical composition to the third order in $\vphi$,
\begin{equation}\label{epsilon_expansion}
\veps(\vphi)=\veps_c+\veps_1\vphi+\frac12\veps_2\vphi^2+\frac16\veps_3\vphi^3.
\end{equation}

The mapping to the language of the Landau expansion Eq. 
(\ref{landau_expansion}) is
\bea
h_c&=&-\frac12 M\veps_c~~~~~~~~~~ h_1=-\frac12 M\veps_1,\nn\\
h_2&=&-\frac12 M\veps_2~~~~~~~~~~ h_3=-\frac12 M\veps_3.\nn
\eea
From Eq. (\ref{tstar_landau}) we find $t^*$ to be
\bea
t^*=\frac18\frac{M^2\veps_3^2}{d}+\frac12 M\veps_2,
\eea
As denoted above, $t^*$ depends on the amplitude of 
the external field, $M$, both linearly and quadratically. 

We can estimate $t^*$ by using the values of $\veps(\phi)$ measured in the 
experiment by Debye and Kleboth \cite{debye_jcp1965}. They found 
$\veps_2=28.6$ and $\veps_3=-116.4$. 
Since $M$ is typically much smaller than unity, the $M^2$ 
term in $t^*$ can be safely neglected and one finds that $t^*\approx 14M$. If 
$M=10^{-3}$ and $T_c=350^\circ$K we find $T^*-T_c\approx 5^\circ$K.

{\bf Nucleation around a charged particle.} Consider a charged spherical
colloidal particle in an ambient gas environment. The particle could be a 
molecular cluster ionized by solar radiation or an aerosol particle lifted from 
the ground. When the particle has a fixed surface charge density $\sigma$, it 
is easy to solve the Laplace's equation to find that $|E|=\sigma 
R_p^2/(\veps(\phi(r))R^2)$, where $\phi$ is the fluid density scaled by 
its critical value, $R_p$ is the particle's radius, and $R$ is 
the distance from its center. The liquid nucleation around the colloid 
is governed by the free energy
\bea
f=f_b(\phi)+\frac12 \frac{M}{\veps(\phi)} r^{-4}.\nn
\eea
Here $M=v_0\sigma^2/(\veps_0k_BT)$ and $r=R/R_p$.

If one truncates the series Eq. (\ref{epsilon_expansion}) at linear order,
$\veps(\phi)=\veps_c+\veps_1\vphi$, then
\bea
h(\vphi)&\simeq&\frac12 
\frac{M}{\veps_c}\left[1-\left(\frac{\veps_1}{\veps_c}\vphi\right)+
\left(\frac{\veps_1}{\veps_c}\vphi\right)^2-\left(\frac{\veps_1}{\veps_c
}\vphi\right)^3\right],\nn\\
\eea
and it is easy to make the following mapping
\bea
h_c&=&\frac12 \frac{M}{\veps_c}~~~~~~~~~~~~ h_1=-\frac12 
M\frac{\veps_1}{\veps_c^2},\nn\\
h_2&=&\frac12 M\frac{\veps_1^2}{\veps_c^3}~~~~~~~~~ h_3=-\frac13 
M\frac{\veps_1^3}{\veps_c^4}.\nn
\eea
The value of $t^*$ is now [Eq. (\ref{tstar_landau})],
\bea
t^*=\frac{1}{18}M^2\frac{\veps_1^6}{d\veps_c^8}-\frac12 
M\frac{\veps_1^2}{\veps_c^3}.
\eea

For binary mixtures of two nonpolar solvents, $\veps_1\sim 1$ 
and $\veps_c\sim 1$. Neglecting $M^2$ compared to $M$, we find that 
$t^*\sim -M$ and $T^*$ is smaller than $T_c$.

{\bf Magnetic spins.} Consider an incompressible monolayer of 
identical spins characterized by a local magnetization areal density $m$ with a 
nearest-neighbor coupling parameter $J$. The spins orient only in the $\pm 
\hat{z}$ direction and the monolayer is
constrained to lie in the $x$--$y$ half-plane with $x<0$. An infinitely long 
wire of radius $R_w$ carrying current $I$ passes through the $y$ axis. The 
magnetic field in the plane is ${\bf B}=\mu_0 I\hat{z}/(2\pi x)$, where 
$\mu_0$ is the vacuum permeability. The coupling term is then $h(m)q(x)=\mu_0 
I m/(2\pi x)$. The governing 
equation for the local magnetization $m$ is given by
\bea
-Jm+\frac12 k_BT\ln\left(\frac{1+m}{1-m}\right)+\mu_0 I/(2\pi x)=0.
\eea
The resulting magnetization is monotonic and smooth in $x$ for small 
currents $I$, and exhibits a sharp coexistence between high and low 
magnetizations in sufficiently large currents. In this simple case, $h'(m)$ is 
independent on $m$ and thus $T^*=T_c$,  but other, nonlinear, forms for $h(m)$ 
can be envisioned and in that case $m(x)$ can also be non-monotonic.

{\bf Laser-tweezing phase separation.} A Laser radiation passing through a 
mixture of two transparent dielectric liquids brings about a dielectrophoretic 
force proportional to the beams intensity $I$. If heat adsorption is
negligible and azimuthal (cylindrical) symmetry applies, the equilibrium 
composition profile is given by 
\bea
f_b'(\phi)-\alpha n(\phi) I(r)-f_b'(\phi_0)=0.
\eea
$n(\phi)$ is the index of refraction and $\alpha$ is a constant 
\cite{wynne_nature_chem2018}. $q(r)=e^{-r^2/r_0^2}$
when the beam is Gaussian, with $r_0$ being the characteristic width of the 
beam. Hence, $T^*=T_c$ if $n$ varies linearly with $\phi$, but $T^*$ can be 
larger or smaller than $T_c$ depending on the sign and magnitude of $n''(\phi)$ 
and $n'''(\phi)$.

{\bf Flow-induced condensation.} Imagine a fluid constrained to lie in an 
infinite plane. The fluid will flow towards the origin if the pressure at the 
orifice of radius $a$ at the origin is smaller than at infinity. In azimuthal 
symmetry and denoting by $r$ the distance from the origin scaled by $a$, the 
fluid's composition in steady-state obeys the equation,
\bea
\phi''+\frac{1}{r}\phi'-f_b'(\phi)+J_0\ln(r)+f_b'(\phi_0)=0.
\eea
Here, primes denote differentiation with respect to the radial coordinate $r$, 
$\phi_0=\phi(r\to\infty)$ and $J_0$ is proportional to the rate of influx at 
$r=1$. At large length-scales, $\phi''$ and $\phi'$ can be neglected and we 
find $h(\phi)=J_0\phi$ and $q(r)=\ln(r)$. Thus $T^*=T_c$ and liquid-gas 
coexistence occurs below $T_c$ if the influx rate $J_0$ is large enough.


\section{Conclusions}

We study a generic mean-field model of systems in external fields. The field may 
have arbitrary dependence on the order parameter $h(\phi)$ and on the spatial 
coordinate $q(r)$. 
This treatment generalizes numerous works on vastly different systems. The 
graphical construction shown in Figs. \ref{fig_graph_sol_f2p}-- 
\ref{fig_finding_phi0} and \ref{fig_g_of_phi} provides 
a simple interpretation of the governing equations Eqs. 
(\ref{3eqs_eq1})-(\ref{3eqs_eq3}) and leads to the phase diagrams in Figs. 
\ref{fig_phase_diagram1} and \ref{fig_phase_diagram2}. Monotonic fields have a 
point in space where they are maximal and this point is analogous to a surface 
in wetting phenomena. We find that the bulk critical point expands to a line of 
second-order phase transition that ends in a ``surface critical point''. A line 
of first-order transition meets the second-order line at this point. This line 
resembles the prewetting line in wetting phenomena \cite{schick_prb1983} and the 
adsorption increases discontinuously on it (not shown) 
\cite{cahn_jcp1977,pgg_rmp1985,evans_mol_phys1983,evans_mol_phys1984, 
schick_prb1984}. 

The second-order line depends on $q(r)$ while the 
first-order line is universal and independent of the form of $q(r)$. 
If some conditions apply to $h(\phi)$, non-monotonic profiles 
can form even when external forcing $\sim q(r)$ is monotonic. It would be 
interesting to investigate how the order-parameter evolves dynamically to produce
the equilibrium profiles found here. We predict an interfacial instability as a 
result of competition between the long-range nature of the force and the local 
diffusive transport and surface-tension related stress. 

The current model applies to isotropic phases but it can be extended to the 
coexistence of isotropic and ordered phases, for example, near the 
disordered-lamellar phase boundary in block-copolymers or similar systems. The 
modulated phases can be generally described by a negative gradient-squared term 
and a higher-order derivative term \cite{andelman_mod_phases,tsori_epl2001}. 
Within the sharp-kink approximation, Eqs. (\ref{3eqs_eq1}--(\ref{3eqs_eq3}) would 
stay the same if the length associated with the modulation period is much smaller 
than the length-scale of the external force. For a strong enough force, the surface 
would induce order (e.g., lamellar) near it even when the bulk is disordered. The 
corresponding profile would have oscillations near the surface, decaying to a uniform 
value at the bulk. Although the model can be extended to modulated phases at least 
in some limits, there are issues related to the direction of the force with respect to 
the crystal axis of the ordered phase. It is not obvious that lamellae would form  
parallel to the lamellar-disorder phase boundary or how an external field couples 
to a body-faced cubic phase.

{\bf Acknowledgment} 

This work was supported by the Israel Science 
Foundation (ISF) grant No. 274/19.

\bibliography{phase_lines_external_force5_cond-mat.bbl}

\end{document}